\documentclass[aps,prl,twocolumn,showpacs,groupedaddress,\if{lengthcheck}\fi
]{revtex4}
\usepackage{amsmath,amssymb,xspace,graphicx}
\renewcommand{\r}{{\bf r}}
\renewcommand{\k}{{\bf k}}
\newcommand{\g}{{\bf g}}
\renewcommand{\v}{{\bf v}}

\begin{document}

\title{Do current-density nonlinearities cut off the glass transition?}

\author{M. E. Cates$^1$}
\author{Sriram Ramaswamy$^2$}
\affiliation{$^1$SUPA, School of Physics, University of Edinburgh, JCMB
Kings Buildings, Edinburgh EH9 3JZ, United Kingdom\\
$^2$Centre for Condensed Matter Theory, Department of Physics, Indian Institute of Science, Bangalore 560012, India, and
CMTU, JNCASR, Bangalore 560064, India}

\date{\today}

\begin{abstract}
Extended mode coupling theories for dense fluids predict that nonlinear current-density couplings cut off the singular `ideal glass transition', present in the standard mode coupling theory where such couplings are ignored. We suggest here that, rather than allowing for activated processes as sometimes supposed, contributions from current-density couplings are always negligible close to a glass transition. We discuss in schematic terms how activated processes can nonetheless cut off the transition, by causing the memory function to become linear in correlators at late times.
\end{abstract}

\maketitle

Mode Coupling Theory (MCT) offers an approximate but appealing account of the glass transition in simple liquids, whose physical content has long been debated.
Judiciously used, particularly in colloidal liquids \cite{pham}, MCT yields experimentally validated, semiquantitative predictions for the slowing down and then arrest of density fluctuations; yet when applied to much simpler systems, such as a single particle in an anharmonic trap, it yields wrong results \cite{sethna}. It is thus important to delineate what the approximations of MCT mean physically, so that these may be improved upon, and/or eventually unified with other theoretical approaches to glass formation.

In this Letter we take this forward by examining the full or `extended' versions of MCT (eMCT) in which nonlinear couplings of the density field to kinetic degrees of freedom (currents) are considered. 
In two important and technically substantial papers, Das and Mazenko \cite{dasmazenko} and subsequently Goetze and Sjoegren \cite{goetzesjoegren} showed that, within both a field-theoretic and a projection-based formulation of eMCT (denoted eMCT1 and eMCT2 respectively), such couplings cut off the dynamical singularity corresponding to the glass transition. 
(Some subsequent variants \cite{schmitz,mazenkoyeo}, denoted eMCT3-5 are addressed below; an up-to-date review is provided by Das \cite{das}.)
This cutoff is not present in the ``standard" MCT (sMCT) which neglects nonlinear coupling to currents \cite{leutheusser,bengtzelius,goetze,ramaswamytoner}.

This cutoff was argued by some to represent an activated hopping mechanism \cite{sjoegren1,goetzereview}. The existence of such a mechanism is physically plausible: the activation energy for relaxation should not exceed that required to expand a volume of order $\xi^3$ to a density low enough to melt the glass. Here $\xi$ is the smallest length large enough that interactions across the boundary of such a melted droplet do not force it to refreeze in the same local state \cite{biroli,rfot}. Since sMCT predicts finite elastic moduli, this energy is finite, and the ideal glass transition is cut off, unless $\xi$ is infinite throughout the glass phase.

In recent years, however, the hopping interpretation of eMCT has seemed increasingly doubtful
(see \cite{das}). From a field-theory viewpoint MCT is formally a re-summed (one-loop selfconsistent) low  temperature expansion \cite{dasmazenko,dasmazenkoreview,miyazaki,schmitz}. In such an expansion, activated terms, which are exponential in $\beta = 1/k_BT$,  represent essential singularities (``instantons'') and should not be accessible even after resummation \cite{bouchaud}. Second, the physics of eMCT should reduce to sMCT for interacting Brownian particles, which conserve neither momentum nor energy \cite{fuchs,miyazaki}. It would be surprising if there were no hopping processes in that case; yet MCT predicts none.  

Below we show that doubts about the physical basis of eMCT are well-founded. Accepting for these purposes that Refs.\cite{dasmazenko,goetzesjoegren} correctly implement their eMCTs, we argue that the resulting cutoffs likely stem from new approximations made in eMCT, {\em beyond} those of sMCT. We show that eMCT forms for the memory function $\gamma_k(t)$, describing retarded friction, violate physical arguments based on a formal separation of current-density terms. 
We then discuss schematically how instantons \cite{rfot,reichman} may nonetheless cut off the sMCT glass transition. 
Our work may thus bring closer a unified description of collective arrest \cite{goetzereview} and activated dynamics \cite{rfot,biroli,kobanderson}.

Our approach closely follows Zaccarelli et al. \cite{zaccarelli,zaccarelli2}. We start with $j = 1...N$ Newtonian particles of mass $m$ interacting via a nonsingular pair potential $v(r)$ in volume $V$, at mean density $\rho_o = N/V$. The density is $\rho_\k(t) = \sum_je^{i\k.\r_j(t)}$ and $(1-\rho_oc_k)^{-1} = S_k = \langle\rho_{-\k}\rho_\k\rangle/N$ is the static structure factor. By standard arguments \cite{zaccarelli}, an equilibrium memory function $\gamma_k(t)$ exists so that \begin{equation}
\ddot\rho_{\k}(t) +\Omega_k^2 \rho_\k(t) +Y_\k(t) = f_{\k}(t)\label{one}\end{equation}
where $Y_\k = \int_0^t\gamma_k(t-t')\dot\rho_\k(t')dt'$; $f_{\k}$ is a noise of mean zero; and $\Omega_k^2 = k^2/\beta mS_k$.
Moreover, one may show from Newton's laws  \cite{zaccarelli} that the memory function obeys
\begin{equation}
\gamma_k(t) = (\beta m/Nk^2)\langle {\cal F}_{-\k}(0)({\cal F}_\k(t) + Y_\k(t))\rangle \label{newtwo}
\end{equation} 
with ${\cal F}_\k = \Omega_k^2\rho_\k+ {\cal T}_\k + {\cal Q}_\k$; ${\cal T}_\k = -\sum_j(\k.\dot\r_j(t))^2e^{i\k.\r_j(t)}$; and $mV{\cal Q}_\k = -\sum_{\k'}v_{k'}(\k.\k')\rho_{\k-\k'}(t)\rho_{\k'}(t)$.
Suppose now that the particle velocities in ${\cal T}$ relax, on some finite timescale $\tau$ that does not diverge near any glass transition (below denoted `rapid' relaxation), towards a global Boltzmann distribution, for which $\langle \dot r_{l\alpha}\dot r_{l\beta}\rangle = \delta_{\alpha\beta}/\beta m$ \cite{zaccarelli}. If so, then for $t \gg \tau$, ${\cal T}_\k(t)$ can safely be replaced by $-k^2\rho_\k(t)/\beta m$. With this replacement (but noting that their Eq.11 fails without it) one recovers in combination Eqs.11 and 12 of \cite{zaccarelli}: that is, $\gamma_k(t) = \gamma_k^s(t)$ where
\begin{eqnarray}
&&\gamma_k^s(t) =(\beta m/Nk^2)[(k^2\rho_oc_k/\beta m)^2 \langle \rho_{-\k}(0)\rho_\k(t)\rangle
\label{newthree}\\ 
&&- (k^2\rho_oc_k/\beta m)\lbrace\langle\rho_{-\k}(0){\cal Q}_\k(t)\rangle + 
\langle{\cal Q}_{-\k}(0)\rho_\k(t)\rangle
\nonumber\\
&&
+\langle\rho_{-\k}(0)Y_\k(t)\rangle\rbrace+\langle{\cal Q}_{-\k}(0)Y_\k(t)\rangle +\langle {\cal Q}_{-\k}(0){\cal Q}_\k(t) \rangle \nonumber
]
\end{eqnarray}
By treating momentum fluctuations as fast and independent of density, $\gamma_k^s(t)$ neglects all the collective fluid modes responsible for current-density couplings; but, for $t\gg \tau$, nothing else of consequence seems neglected from (\ref{newthree}) as derived in Refs.\cite{zaccarelli,zaccarelli2}. As shown there, $\gamma_k^s(t)$ further reduces to the well known sMCT memory function $\gamma_k^{sMCT}(t)$, if one treats $\rho_\k(t)$ as a Gaussian random process, and assumes that $\zeta \equiv\beta v_k + c_k = 0$. We discuss such steps below, but {\em without making either of them} write
\begin{equation}
\gamma_k(t) = \gamma^s_k(t) + \Delta\gamma_k(t)
\label{two}
\end{equation}
where all ``current-density couplings'' reside solely in the second term. 
By this we mean that, although $\gamma^s_k(t)$ still formally depends on $\dot\rho$ (via $Y(t)$), it describes a system in which
currents and densities are explicitly {\em decoupled}, in the sense that {\em instantaneous equipartition} holds for all averages of equal-time products of particle velocity pairs in (\ref{newtwo}).
The coupling term, $\Delta\gamma_k(t)$, whose explicit form follows by subtracting (\ref{newthree}) from (\ref{newtwo}), then contains various correlators involving $\Delta{\cal T}_\k = {\cal T}_\k + k^2\rho_\k/\beta m$. In real space this is $\Delta{\cal T}(\r,t) =
\nabla_{\alpha} \nabla_{\beta} \sum_l 
[\dot{r_{l \alpha}} \dot{r_{l \beta}} - \delta_{\alpha\beta}/\beta m] \delta(\r-\r_l(t))$.

Remarkably, both the eMCT1 of \cite{dasmazenko}
and the eMCT2 of \cite{goetzesjoegren}
yield approximations for the memory function that depart from the additive structure shown in (\ref{two}). In the relevant regime (at low frequencies $\omega$,  with $k$ around the peak in $S_k$, and  in the glassy
regime where $\gamma_k(t)$ decays very slowly), the results are \cite{dasmazenko,goetzesjoegren,footderive}
\begin{eqnarray}
\gamma^{eMCT1}_k(\omega) &=& (1+iA_k/\omega)\gamma^{sMCT}_k(\omega)\label{four}
\\
\gamma^{eMCT2}_k(\omega) &=& \frac{\gamma^{sMCT}_k(\omega)}{1-iB_k\gamma^{sMCT}_k(\omega)}\label{fourb}
\end{eqnarray}
Such results seemingly require that at low frequency the kinetic correction $\Delta\gamma_k(\omega)$ either  becomes larger than the standard contribution $\gamma_k^s(\omega)$ (in the case of (\ref{four})), or remains comparable to it (in the case of (\ref{fourb})). 
We now argue that such behavior cannot be a physical consequence of the current-density couplings as defined above, because $\Delta\gamma_k(t)$ must remain negligible near a glass transition,
even when the velocity and temperature fields of the fluid (the only significant physics omitted from $\gamma_k^s(t)$), are accounted for.

To see this, we define $\rho(\r,t), \v(\r,t)$ and $T(\r,t) = T + \delta T$ as hydrodynamic number density, velocity and temperature fields, formed by local averaging
over regions $\delta V(\r)$ of volume $\lambda^3$ containing  $\nu(\r,t)>1$ particles \cite{landau}. The collective velocity field is
$\v(\r) =\sum_{j\in\delta V(\r)}\dot\r_j/\nu(\r)$ and 
the density field is $\rho(\r) = \nu(\r)/\lambda^3$. 
Setting $\dot\r_l(t) -\v(\r_l,t) = \delta \v_l(t)$, we next insist that relaxation of the {\em peculiar} velocities $\delta\v_l(t)$ is rapid: 
{\em local equilibrium alone} allows the replacement
$
m\langle \delta v_{l \alpha}(t) \delta v_{l \beta}(t)\rangle \simeq (1-1/\nu(\r,t)) 
\delta_{\alpha \beta} k_BT(\r_l,t)
\nonumber
%\label{s1}
$  
\cite{footnu}.
At temporal and spatial scales relevant to collective modes,  we can then substitute in (\ref{newtwo}) 
${\cal T}(\r,t) \simeq$
$\nabla_\alpha\nabla_\beta[\rho(\r,t) v_\alpha(\r,t) v_\beta(\r,t)+ (\rho(\r,t)-\lambda^{-3}) k_BT(\r,t)/m]$. 
This in turn can be written $\sum_{i=1}^3{\cal T}_i(\r,t)$, where
$m{\cal T}_1 =  
m\nabla_{\alpha} \nabla_{\beta}( 
\rho v_{\alpha} v_{\beta})$;  
$m{\cal T}_2 =  
k_BT \nabla^2 \rho$; and 
$m{\cal T}_3 =k_B[\delta T\nabla^2\rho+(\rho-\lambda^{-3})\nabla^2\delta T+ (\nabla\delta T)\nabla \rho]$.
These describe respectively gradients of hydrodynamic inertial stress; gradients of ideal gas kinetic pressure; and a non-isothermal correction to the latter.

We finally argue that heat conduction is rapid ($\delta T \simeq 0$) in the relevant $k$-range. 
This isothermal approximation 
 is standard, even within eMCT \cite{dasmazenko,goetzesjoegren,schmitz,mazenkoyeo,das},
although it fails at lower $k$, where ${\cal T}_3$ causes sound-speed renormalization 
\cite{footsoundspeed}. 
Using it, we find that $\Delta {\cal T}(\r,t)$ is merely ${\cal T}_1(\r,t) = \nabla_\alpha\nabla_\beta \rho v_\alpha v_\beta$, so that $\Delta\gamma_k(t)$ contains only correlators of ${\cal T}_1(\r,t)$ with $\nabla^2\rho(\r,t)$; with ${\cal Q}(\r,t)$; and with itself.
The first of these terms, $\langle \nabla^2\rho({\bf 0},0)\nabla_\alpha\nabla_\beta\rho v_\alpha v_\beta(\r,t)\rangle$ in real space, cross-correlates gradients of density and inertial stress.
But regardless of $\nabla^2\rho({\bf 0},0)$, the inertial stress term $\rho v_\alpha v_\beta(\r,t) \equiv \sum_{i,j\in\delta V(\r)}\dot r_{i\alpha}\dot r_{j\beta}/\nu(\r)\lambda^3$ rapidly decays (on time scale $\tau_1$, say) to its equipartition value $k_BT\delta_{\alpha\beta}/m\lambda^3$; this is $\rho$-independent \cite{landau}, hence uniform, and killed off by $\nabla_\alpha \nabla_\beta$. Accordingly $\langle \nabla^2\rho({\bf 0},0){\cal T}_1(\r,t)\rangle \simeq 0$ for $t\gg \tau_1$. The only alternative is if relaxation of $\rho v_\alpha v_\alpha$ is not rapid, but remains {contingent} on that of $\rho$ itself, causing $\tau_1$ to become very large in the glassy regime. Remarkably, a mode-coupling ansatz 
$\langle \nabla^2\rho({\bf 0},0){\cal T}_1(\r,t)\rangle  \simeq \langle\nabla^2\rho({\bf 0},0)\nabla^2\rho(\r,t)\rangle \langle v^2(\r,t)\rangle$ says exactly this. But such a result is wrong: it implies that, e.g., in window-glass, where atomic {\em positions} remain unrelaxed for centuries, the atomic {\em momenta} 
(of which $\v$ is a local average) fail to reach equipartition, even on that time scale.
The same reasoning gives $\langle{\cal T}_1({\bf 0},0){\cal Q}(\r,t)\rangle\simeq 0$. It also gives $\langle{\cal T}_1({\bf 0},0){\cal T}_{1}(\r,t)\rangle\simeq 0$, with one caveat;
this contains a product of squares of two velocities at unequal times. 
It might therefore show algebraic, rather than exponential, decay \cite{ernstltt}. But such long-time-tails should play no role in the physics of glasses. (Indeed they are explicitly neglected in eMCT2 \cite{goetzesjoegren}.)
  
The above arguments complete our case that $\Delta\gamma_k(t)$ in (\ref{two}), 
containing formally all current-density couplings, can be safely neglected anywhere near a glass transition. The results (\ref{four},\ref{fourb}), which seemingly require $\Delta\gamma_k(t)$ to become large, must then result from one of two things. Either these theories make injudicious approximations to $\Delta\gamma_k(t)$ and should be rejected; or they deal with the averages in (\ref{newthree}) in a different manner from sMCT, so that (\ref{four},\ref{fourb}) stem not from current-density couplings (as defined above) but from different handling of the ``standard'' piece of the memory function, $\gamma_k^s(t)$. Recall that to reduce $\gamma_k^s(t)$ to $\gamma_k^{sMCT}(t)$, two further approximations are made in \cite{zaccarelli}. The first is to set $\zeta = 0$, which amounts to adopting dynamically the Ramakrishnan Yussouff (RY) free energy functional \cite{rk,das}. The second
is that  $\rho_{\k\neq {\bf 0}}(t)$ may be ``{\em tr}eated {\em a}s {\em g}aussian'' (a practice we below call 
``tragging'') in the specific sense that its three-point correlators are set to zero, and its four-point correlators factored into sums of products of two-point correlators. A change to either assumption might give a cutoff, in principle.

We argue next that eMCT1 stems from injudicious approximation to $\Delta\gamma_k(t)$ rather than from an altered treatment of $\gamma^s_k(t)$. This is because eMCT1 trags not only $\rho_\k$, but also $\v_\k$ and the momentum density $\g_\k$: indeed it aims to determine selfconsistently the covariance matrix of these tragged variables by an optimized resummation of nonlinearities. But if one believes RY is adequate \cite{das}, and $\Delta\gamma_k(t)$ is negligible, then the tragging of the density is {\em sufficient} to show via \cite{zaccarelli} that $\gamma_k(t) = \gamma_k^{sMCT}(t)$.
If so, then since the density {\em remains} tragged in eMCT1 (tragging additional variables does not untrag existing ones), this theory can only give results different from sMCT by making {\em additional} 
approximations that sMCT {\em avoids}.

Let us now consider how (\ref{four}) actually arises. Within the formalism of Ref.\cite{dasmazenko} the renormalized equation of motions are 
$\sum_y G^{-1}_{\hat x y}\,y = f_x$, where the fluctuating (and tragged) variables $y$ comprise $\rho,\v$ and $\g$. $G_{\hat x y}$ is the propagator calculated in Ref.\cite{dasmazenko}. The hatted variables are auxiliary fields used to treat the noise terms $f_x$ (with variances set by $G^{-1}_{\hat x \hat x}$, and cross-correlations by $G^{-1}_{\hat x \hat y}$), and also to treat a constraint described below. 
Omitting noise terms for clarity, eMCT1's equations of motion are \cite{dasmazenko}
\begin{eqnarray} 
-i\omega\rho_\k(\omega) +i\k.\g_\k(\omega) = 0
\label{hydro1}\\
-i\omega\g_\k(\omega) -i\k c^2\rho_\k(\omega)+\eta_k(\omega) k^2\v_\k(\omega) = {\bf 0} 
\label{hydro2}
\\
\g_\k(\omega) - \rho_o\v_\k(\omega) + i\alpha\k\rho_\k(\omega) = {\bf 0}\label{hydro3}
\end{eqnarray}
where $\alpha = A_k/k^2$ is a constant, and we have set $m=1$ so that $\rho$ is now the mass density. 
These are linearized effective equations arising from a complicated nonlinear model; nonetheless one would normally expect them to have the usual hydrodynamic form, with renormalized coefficients. In this sense, Eqs. (\ref{hydro1},\ref{hydro2}) are the expected equations of motion (with $c$ a sound speed, which is $k$-independent in eMCT1); but Eq.(\ref{hydro3}) is not as expected. To understand it, note that in Ref.\cite{dasmazenko}, a field $\hat\v$ is deployed to maintain, as a nonlinear constraint, the defining equation $\g(\r,t)=\rho(\r,t)\v(\r,t)$ of the fluid velocity $\v$. Despite this,
the bare constraint $\g = \rho_o\v$, instead of renormalizing into the nonlinear one, $\g = \rho\v$, becomes (\ref{hydro3}), which in real space reads $\g =\rho_o\v-\alpha\nabla\rho$. Combining with (\ref{hydro1}) (i.e. $\dot\rho = -\nabla.\g$) we obtain
\begin{equation}
\dot\rho = - \rho_o\nabla.\v + \alpha \nabla^2\rho 
\end{equation} 
instead of the continuity equation $\dot\rho = -\nabla.(\rho\v)$.  
Eliminating $\g,\v$ from (\ref{hydro1}--\ref{hydro3}), one obtains
\begin{equation}
-\omega^2\rho_\k + c^2\rho_\k -i(\omega k^2/\rho_o)\eta_k(\omega)[1+iA_k/\omega]\rho_\k = f_\k \label{confirm}
\end{equation}
which confirms (\ref{four}).
Note we find no algebraic error in Ref.\cite{dasmazenko}. The fault appears intrinsic to the whole program of extending MCT, which neglects vertex corrections \cite{miyazaki}, to currents. Here, instead of giving the vertex correction required to convert $\g=\rho_o\v$ into $\g=\rho\v$, eMCT1 creates a new (self-energy) term in $\alpha$ which, at least at the level of effective equations of motion,
violates the constraint that $\hat \v$ was supposed to enforce. The result ({\ref{hydro3}) thereby introduces a parallel diffusion channel
which requires, physically, a momentum sink. 
While it might be tempting to associate the resulting `motion' with activated hopping \cite{goetzesjoegren}, 
the end result (\ref{four}) of eMCT1 is incompatible with the physical content of (\ref{two}) above, and we reject it.

An alternative to eMCT1 (call this eMCT3) in Ref.\cite{schmitz} likewise violates continuity; the authors  `justify' this on the grounds that eMCT3 {\em also} violates Galilean invariance. (An adequate theory should do neither.) The cutoffs predicted by eMCT1 and eMCT3 are almost identical \cite{das2,das}. Ref.\cite{mazenkoyeo} offers two further variants. In eMCT4, the constraint differs by a Jacobian from eMCT1; the results are unaffected. In eMCT5, the constraint is replaced by a series expansion in $\delta \rho(\r)$. The resulting equation of motion for $\g$ is that found by substituting (\ref{hydro3}) in (\ref{hydro2}). Thus, so long as (\ref{hydro2}) is true, (\ref{hydro3}) is implied by eMCT5: the expansion in $\delta\rho$ breaks the constraint $\g=\rho\v$ in exactly the same sense as eMCT1, with the same consequences.

\if{
Such failure might also be a factor in causing the projection-based eMCT2 of Ref.\cite{goetzesjoegren} to give (\ref{fourb}).
For instance, the continuity equation in the form $\dot\rho = -\nabla.\g$ might break down if the density-current self-energy contribution in eMCT2 (denoted $\Gamma^{(2)}_{0,1}(q,z)$ in \cite{goetzesjoegren}) were nonzero. The expression for this  is complicated \cite{sjoellandersjoegren}; we could not confirm whether it is zero or not.
}\fi

We now turn to eMCT2. Here tragged variables are less easily identified \cite{kawasaki,sjoellandersjoegren,goetzesjoegren}, but an authoritative review \cite{gs95}
seems to suggest that the {phase space} density $\tilde\rho_\k({\bf p},t)$ (with ${\bf p}$ particle momentum) is tragged. If so, since $\rho_\k(t) = \sum_{\bf p}\tilde\rho_\k({\bf p},t)$, and any sum of tragged variables is {\em de facto} tragged, $\rho_\k$ is tragged also in eMCT2. If this is true, then the cutoff
(\ref{fourb}) must either: (a) like (\ref{four}), stem from injudicious treatment of $\Delta\gamma_k(t)$ that sMCT bypasses; or (b) stem from avoiding use of $\zeta = 0 $ within the standard part of the memory function, $\gamma_k^s(t)$. In case (b), eMCT2 should be equivalent to Eq.13 of \cite{zaccarelli}, found by tragging the density in (\ref{newthree}): 
this looks highly unlikely. The only remaining case is (c): that tragging of the density is avoided, or partially avoided, in eMCT2. Result (\ref{fourb}) might then remain a physical one; but, at least by our definition, one unrelated to current-density couplings.

Our own view is that $\Delta\gamma_k(t)$ is always negligible; RY (hence $\zeta = 0$) generally sufficient; and the cutoffs in all forms of eMCT likely  approximation-induced. Suppose one accepts this case: by what physical mechanism can the glass transition instead be cut off? Our results show that any cutoff mechanism must {\em specifically invalidate} the tragging of the density. We argue now that instanton-like activated processes (`hopping') do exactly this.

When hopping occurs in a system that would otherwise be fully arrested, the generic relaxation process seen by any local neighborhood is as follows. First, wait a long, stochastic time interval; then a locally large and rapid change occurs which reconfigures the density field nonperturbatively (an instanton). This is very unlike a Gaussian process, in which relaxations occurs in temporally uniform small increments; tragging is appropriate only for relaxation by such gradual events. 
In the limit of pure instanton relaxation, the product of two fluctuating densities, $\rho({\bf 0},0)\rho(\r,t)$, becomes suddenly randomized after a long 
period of being constant; and crucially, 
$\rho^2({\bf 0},0)\rho^2(\r,t)$ decorrelates at exactly the same instant. 
Under these conditions, the (normalized) four-point correlator $\langle\rho^2({\bf 0},0)\rho^2(\r,t)\rangle$, which is almost but not quite the one needed in (\ref{newthree}), is equal to the two-point correlator, not its square, in maximal violation of tragging. 

This suggests a schematic route by which sMCT might be cut off by hopping. (For more complete approaches along similar lines, see \cite{bouchaud,rfot,reichman}.) For $t\ll \tau_I$ (an instanton time) the minimal sMCT schematic model, called $F_{12}$, is applied. This has a single correlator $\Phi(t)$ with memory function
$m(t) = \lambda_1\Phi(t) + \lambda_2\Phi(t)^2$ and $\lambda_{1,2} \ge 0$. On increasing $\lambda_1$ this has \cite{goetzesjoegren}, for $\lambda_2<1$, an unrealistic `type A' transition 
at $\lambda_1^c = 1$; for $1<\lambda_2<4$, a realistic transition at $\lambda_1^c = 2\lambda_2^{1/2}-\lambda_2$; and for $\lambda_2 > 4$,
the system is always in the glass.
Suppose we generalise this to
\begin{equation}\label{schema}
m(t) = \lambda_1(u)\Phi(t) + \lambda_2(u)\Phi(t)^2
\end{equation}
with $u = t/\tau_I$.  This defines parametrically some curve on the $\lambda_1,\lambda_2$ plane; if this starts in the glass phase and ends in the fluid, one can expect behavior appropriate to a cut-off glass (with details dependent on the curve chosen, and on the form of $\lambda_1(u)$).
However, treating the argument of the preceding paragraph literally would lead not only to $\lambda_2(\infty) = 0$, but also to $\lambda_1(\infty) = \lambda_1(0) + \lambda_2(0)$; this cannot give a trajectory of the required character \cite{goetzesjoegren}. This objection is perhaps not fatal since, in the `scheme' that links $F_{12}$ with fully $\k$-dependent sMCT, the $\lambda$'s are proxies for certain $\k$-integrals, whose crossover algebra in the instanton regime need not be as simple as assumed above. We leave detailed study of (\ref{schema}), and its possible relation to non-schematic models, for future work.

On nearing completion of this work, a preprint \cite{biroli2} appeared giving more grounds to doubt the eMCT cutoff.

We thank R. Adhikari, M. Fuchs, M. Greenall, D. Reichman and T. Voigtmann for useful discussions. Work funded in part by EPSRC GR/S10377. 
The Centre for Condensed Matter Theory is funded by the DST, India. Each author thanks the other's institution for hospitality.  
%
%
%%%%%%%%%%%%%%%%%%%%%%%%%%%%%%%%%%%%%%%%%%%%%%%%
%% BACKMATTER
%%%%%%%%%%%%%%%%%%%%%%%%%%%%%%%%%%%%%%%%%%%%%%%%
%
%\vspace{-.6cm}
%
%
%\vspace{-.4cm}

%
%
%

\begin{thebibliography}{1}
\bibitem{footderive}
Eq.(\ref{four}) is rederived below in (\ref{confirm}). Eq.(\ref{fourb}) is (2.21a) of \cite{goetzesjoegren}.
\bibitem{footsoundspeed}
$\Delta\gamma_k(t)$ contains via $\langle{\cal T}_3{\cal T}_3\rangle$ a term in $\langle\delta T_\k(0)\delta T_{-\k}(t)\rangle$. For low-$k$ phonons (now choosing $\lambda^3\gg 1/\rho_o$) this can be replaced by its equal time value, giving a sound speed renormalization $\Omega_k^2 \to\Omega_k^2+k^2 k_B^2T/mc_v$ with $c_v$ the heat capacity per particle.
\end{thebibliography}
\end{document}